\documentclass[12pt]{article}
\usepackage{times}
\usepackage{geometry}
\usepackage[utf8]{inputenc}
\usepackage{enumitem,amssymb}
\usepackage{ragged2e}
\usepackage[dvipsnames]{xcolor}
\newlist{thematic}{itemize}{8}
\setlist[thematic]{label=$\square$}
\usepackage{pifont,caption}

\newcommand{\arcsec}{\mbox{$^{\prime \prime}$}}
\date{\vspace{-10ex}}

\usepackage[utf8]{inputenc}
\usepackage{amsmath,amssymb,wasysym}
\usepackage{graphicx}
\usepackage{sidecap}
\usepackage{titlesec}
\titlespacing*{\section}{0pt}{1ex}{0ex}
\titlespacing*{\subsection}{0pt}{1ex}{0ex}
\titleformat*{\section}{\large\bfseries}
\titleformat*{\subsection}{\normalsize\bfseries}

\title{{\bf Astro2020 Science White Paper:} \\
Spatially Resolved UV Nebular Diagnostics in Star-Forming Galaxies}

\usepackage{natbib}
\usepackage{graphicx}
\newcommand{\emilcheckbox}{\ensuremath{\leavevmode\rlap{$\checkmark$}\square}}

\newcommand{\ii}{{\sc ii}}
\newcommand{\iii}{{\sc iii}}
\newcommand{\iv}{{\sc iv}}
\newcommand{\W}{$\lambda$}

\begin{document}

\maketitle
\begin{center}
{\bf Thematic Areas:} 
\emilcheckbox\enskip Galaxy Evolution\quad 
\end{center}

\begin{center}
{\bf Principal Authors:}\\
Bethan James$^1$ and Danielle Berg$^2$ \\
\smallskip
\footnotesize{
$^1${Space Telescope Science Institute, 3700 San Martin Drive, MD, 21218; bjames@stsci.edu} \\
$^2${The Ohio State University, 140 W. 18th Avenue, Columbus, OH 43202; berg.249@osu.edu} }

\smallskip

\normalsize{{\bf Co-Authors:}\\
Rongmon Bordoloi$^3$,
Nell Byler$^4$, 
John Chisholm$^5$,
Dawn Erb$^6$, 
Nimish Hathi$^1$,
Matthew Hayes$^7$,
Alaina Henry$^1$, 
Anne Jaskot$^8$, 
Lisa Kewley$^4$, 
Sally Oey$^9$,
Molly Peeples$^1$,
Swara Ravindranath$^1$, 
Jane Rigby$^{10}$,
Claudia Scarlata$^{11}$, 
Daniel Stark$^{12}$, 
Jason Tumlinson$^1$,
Peter Zeidler$^{13,1}$}
\\
\smallskip
\footnotesize{
$^3${North Carolina State University}
$^4${Australian National University}
$^5${University of California Santa Cruz}
$^6${University of Wisconsin-Milwaukee}
$^7${Stockholm University}
$^8${University of Massachusetts}
$^9${University of Michigan}
$^{10}${NASA Goddard Space Flight Center}
$^{11}${University of Minnesota}
$^{12}${University of Arizona}
$^{12}${JHU}}
\end{center}

\begin{abstract}
  Diagnosing the physical and chemical conditions within star-forming galaxies (SFGs) is of paramount importance to understanding key components of galaxy formation and evolution: star-formation, gas enrichment, outflows, and accretion. Well established optical emission-line diagnostics used to discern such properties (i.e., metal content, density, strength/shape of ionizing radiation) will be observationally inaccessible for the earliest galaxies, emphasizing the need for robust, reliable interstellar medium (ISM) diagnostics at ultraviolet (UV) wavelengths.
  Calibrating these UV diagnostics requires a comprehensive comparison of the UV and optical emission lines in nearby SFGs. 
  Optical integral field unit (IFU) surveys have revealed the inhomogeneous nature of the ISM in SFGs,
  which leads to non-systematic biases in the interpretation of unresolved sources.
  Spatial variations are especially important to consider at UV wavelengths, 
  where the strongest emission features originate from only the highest excitation regions of the nebula
  and are challenging to distinguish from competing high-ionization sources (e.g., shocks, AGN, etc.).
Since surveys collecting large-scale optical integral field unit (IFU) spectroscopy are already underway,
this white paper calls for an IFU or multi-object far-UV (FUV) spectroscopic instrument with high sensitivity, high spatial resolution, and large field of view (FoV). 
Given the impact of large-scale optical IFU surveys over the past decade, this white paper emphasizes the scientific need for a comparable
foundation of spatially-resolved far-UV spectroscopy survey of nearby galaxies that will lay the foundation of diagnostics critical to the interpretation of the distant universe.  
\end{abstract}

\clearpage

\section{Motivation}
  In this white paper, we argue that securing the local UV scale of physical properties and abundances will provide the foundation for interpreting future observed-frame optical and near-IR observations being enabled by the ELTs (including JWST). In turn, secure determinations of the physical properties of galaxies are an essential prior to many key science questions: To what level are high-z galaxies chemically evolved? By understanding the metal content of gas, we unlock signatures of exchange between SF, gas accretion, and supernova-driven feedback across cosmic time. How do the harsher environments of high-z galaxies affect star-formation/metal retention/escape of ionizing radiation? We know that galaxies at high-z have higher SFR, higher electron densities, and more extreme ionizing fields, but we are yet to accurately quantify to what degree and fully explore their effects. Why are extreme nebular emission lines prevalent at high-z, and what mechanisms promote Lyman Continuum escape? With our currently limited ability to map the ionization structure of the ISM (see Figures~1 and 2), our over-simplified view of the ISM prevents us from sampling the wide range of physical processes that are clearly ongoing in SFGs. 
    
\textit{We need to understand all of these properties in concert in order to determine what is driving differences in populations of SFGs at different redshifts. This, in turn, has important implications for the application / interpretation of metallicity calibrations, dust corrections, stellar populations, feedback and metal retention, and more. As such, it is imperative that we obtain spatially resolved rest-frame UV spectroscopy across nearby SFGs to establish the essential diagnostic tools needed to understand fundamental questions concerning the evolution, physical conditions, and ionization structure of SFGs across cosmic time.}

\section{Understanding SFGs: The Power of Emission-Line Diagnostics}
For over six decades we have used optical (and IR) emission lines to unravel and decipher the complex conditions within the ionized gas of SFGs. Empirical relationships between emission line strengths and the physical conditions of the gas from which they originate have been readily exploited to provide a suite of properties such as temperature, density, metal content, and shape and strength of the ionizing radiation field (see Maiolino \& Mannucci 2019 for a review). Moreover, with the application of detailed photoionization models, we expanded our diagnostic tool-set by calibrating nebular line ratios with additional parameters such as radiation field hardness (O$_{32}$ -- U-parameter, e.g., D\'{i}az+2000; Kewley \& Dopita 2002; Kewley et al. 2013), metallicity-indicators (R$_{23}$ -- $Z$, e.g.,  Kobulnicky \& Kewley 2004; Pilyugin \& Thuan 2005; Nagao+2006, Curti+2017), or the source of ionizing radiation (e.g., the BPT diagram: [O~\iii]/H$\beta$ vs [N~\ii]/H$\alpha$, Baldwin+81; Kewley+2001; Kauffmann+2003; Lamareille 2010). As such, we now have the ability to predict a wide range of properties using optical strong emission-lines (SLs) readily available in even the faintest galaxies. \looseness=-2

Without these diagnostics, we remain unaware of several key galaxy properties that regulate the formation and evolution of galaxies such as star-formation/feedback. 
The chemical evolution of SFGs is a key tracer of holistic galaxy evolution, as the ISM encapsulates the interplay between SF, gas accretion, and outflows across cosmic time.
To this end, we can use emission line diagnostics to both determine how these processes evolve across cosmic time and over a large volume of galaxies (i.e., galaxy surveys such as MaNGA). 
As a result, relationships  such as the unique interplay between metal-content and star-formation have emerged (Barrera-Ballesteros+2018).

\begin{figure}
  \label{fig:BPT}
  \centering
  \includegraphics[width=1\textwidth]{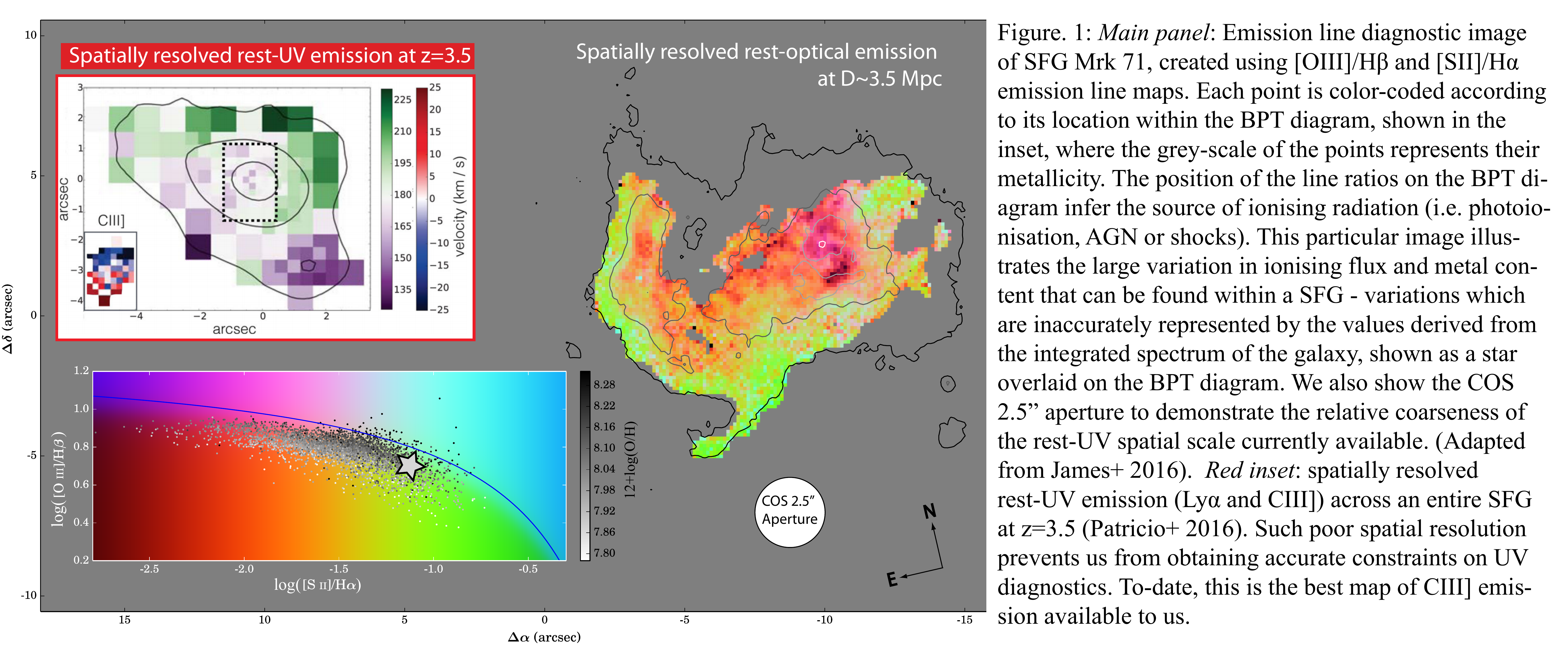}
\vspace{-2ex}
\end{figure}

\section{The 3D Spectroscopic Revolution}
Over the past two decades, the advent of integral field units (IFUs) has revolutionized our ability to spectroscopically probe the spatial distribution of nebular physical conditions.  
IFU spectroscopy has allowed us to map sensitive emission lines across galaxies, both near (e.g., Belfiore+2016; Davies+2017; Micheva+2019) and far (e.g., Newman+2014), with ever increasing spatial resolution, sensitivity, and FoV. In turn, the topography of nebular line ratios has been explored across nearby galaxies, spawning a series of important discoveries regarding emission-line diagnostics:

\begin{itemize}[leftmargin=*]
    \item Mapping both the auroral and SL metallicity determinations have shown that galaxies are not always chemically homogeneous. It is a well known fact that massive, spiral galaxies have metallicity gradients across their disks (e.g., \'{S}anchez+2014, Ho+2015, Berg+2015). However, even observations of individual H~\ii\ regions show variations, including temperature fluctuations (e.g., Peimbert 1967) and chemical inhomogeneities that highlight regions of enrichment (e.g., James+2009, James+2013a, Westmoquette+2013, Kumari+2018) and/or metal-poor gas inflows (e.g. James+2013b, Kumari+2017). 
    Most importantly, however, 
    \textit{integrated light spectra do not accurately represent the spatially-resolved average metallicity in these regions} (e.g. Figure~1), and can be further biased by diffuse ionized gas (DIG, e.g., Sanders+2017, Wielbacher+18, Bik+18), resulting in unreliable SL diagnostics.
    \item Mapping diagnostic line ratios (e.g., BPT diagrams, Figure~1) has allowed us to isolate and disentangle different sources of ionizing radiation, such as AGN/shock fronts (e.g., Groves+2008, Davies+2017, Rich+2011, D'Agostino+2019). In turn, we have identified that the contribution from shock-ionized gas in SL diagnostics is non-negligible and that feedback mechanisms of this kind can suppress star-formation.
    \item Mapping the strength of the ionizing radiation (e.g. O$_{32}$ line ratio) shows significant variation throughout SFGs (e.g., Fensch+2015, Poetrodjojo+2018). In many cases, such maps have enabled us to discern between ionization- versus density-bound regions (i.e., essentially imaging the Stromgren sphere), and reveal the mechanisms that allow for the escape of Lyman continuum photons (e.g. Zastrow+2011, Micheva+2018, Keenan+2017). Such detailed understanding of the ionization structure within galaxies is simply not achievable via integrated light measurements. Further, we have yet to understand the effects of electron density in such systems as the strongest optical density diagnostic line ratios can only probe the low-density regime ($n_e\sim100$~cm$^{-3}$) towards the outer region of the nebula.
\end{itemize}

\section{Diagnostics at High-$z$: enlisting the UV}
While the advent of optical IFU studies has significantly advanced our understanding of galaxy evolution,
we are on the precipice of a large problem. As we enter the {\it JWST} and ELT era, the emission lines that we will access from the most distant galaxies will be in the rest-frame UV.
The UV contains a suite of emission lines 
(see Figure~2) that can be used to characterize a plethora of gas properties including temperature, density, and metal content (both nebular+ISM, e.g., James+14, Erb+2010), as well as reflect the shape and hardness of the ionizing spectrum (Berg+2019). 
The photon collecting power of forthcoming ELTs and {\it JWST}, combined with the visibility of these UV lines in $z>5$ galaxies with standard rest-frame IR bands (shown in Figure~3), will open an unprecedented window
onto reionization-era systems. 
As such, it is imperative that we understand and calibrate the UV emission line diagnostics 
that will allow us to bridge the local and high$-z$ universe.

\begin{itemize}
\item The metallicity of a galaxy is typically characterized by its oxygen abundance, which, for the past $\sim$50 years, has been widely determined from optical emission lines for nearby galaxies (e.g., Searle 1971).
Several efforts have been successful in using FUV emission lines to probe the metal content of the ionized ISM in SFGs (Du+17; Stark+17), and the outlook is promising for rest-UV emission features as accurate probes of the metal content of high-$z$ galaxies. If developed, reliable UV metallicity indicators would allow the first uniform assessment of chemical enrichment across redshift. 
C~\iii] emission is especially strong in metal-poor galaxies (e.g., Rigby+15; Senchyna+17), and is proving to be a powerful diagnostic at high-$z$, showing correlations with the temperature-sensitive O~\iii] \W1666 and [O~\iii] \W4363 lines (Jaskot \& Ravindranath 2016) and Ly$\alpha$ EW (e.g., Shapley+2003, Bayliss+14). However the relationship between EW(C~\iii]) and metallicity is yet to be well constrained at high-$z$, especially in the low-metallicity regime where a large scatter remains (see Fig.~3, Byler+2019, Nakajima+18).  Alternatively, Byler+2018 used photoionization models to show that O~\iii] \W\W1660,1666 / Si~\iii] \W\W1883,1892 could be a useful metallicity indicator. 
However, we still need to empirically test the utility of these abundance indicators on a spatially resolved basis. High S/N rest-frame UV spectra from H~\ii\ regions with varying physical properties across galaxies are needed to determine the parameter space over which the UV metallicity indicators are appropriate. \looseness=-2

\item The rest-frame UV offers two sets of SL doublets whose ratios are sensitive to the electron density, $n_e$, of the ionized gas: C~\iii] \W1907/\W1909 and Si~\iii] \W1883/\W1892. 
Several cases exist where the doublet pairs have been resolved, at both the low- and high-$z$ regime, and used to derive an $n_e$ (e.g., James+14; James+18;  Christensen+12; Bayliss+14; Berg+18; Berg+19). The emissivity ratios of these two density diagnostics demonstrate that they do probe regions of higher $n_e$ compared to their optical counterparts. 
However, cases exist where the UV doublets reveal far higher $n_e$ than derived from density diagnostics in the optical low-ionization regime, and ratios that are unphysical (e.g., Berg+19) suggesting that the mechanisms producing these doublets are not yet fully understood. 
While $n_e$ is perhaps somewhat overlooked as an important parameter in the grand scheme of galaxy evolution,
`high' electron densities have been linked to mechanisms that both aid Lyman continuum escape (due to a clumpy medium) and prevent the recycling of metals due to the suppression of superwinds (e.g., Jaskot+17, Jaskot+19), 
and have also been proposed as being responsible for the `BPT offset' between low- and high-$z$ galaxies (e.g., Sanders+16).
As such, it is imperative that we understand how these $n_e$ diagnostics behave and why. This can only be achieved with spatially resolved data that covers a wide range of electron densities such as those seen in the local universe ($10^2 < n_e\ {\rm cm}^{-3} < 10^6$) and a range of chemical and physical conditions.

\begin{SCfigure}
  \includegraphics[scale=0.5,trim=0mm 0mm 0mm 0mm,clip]{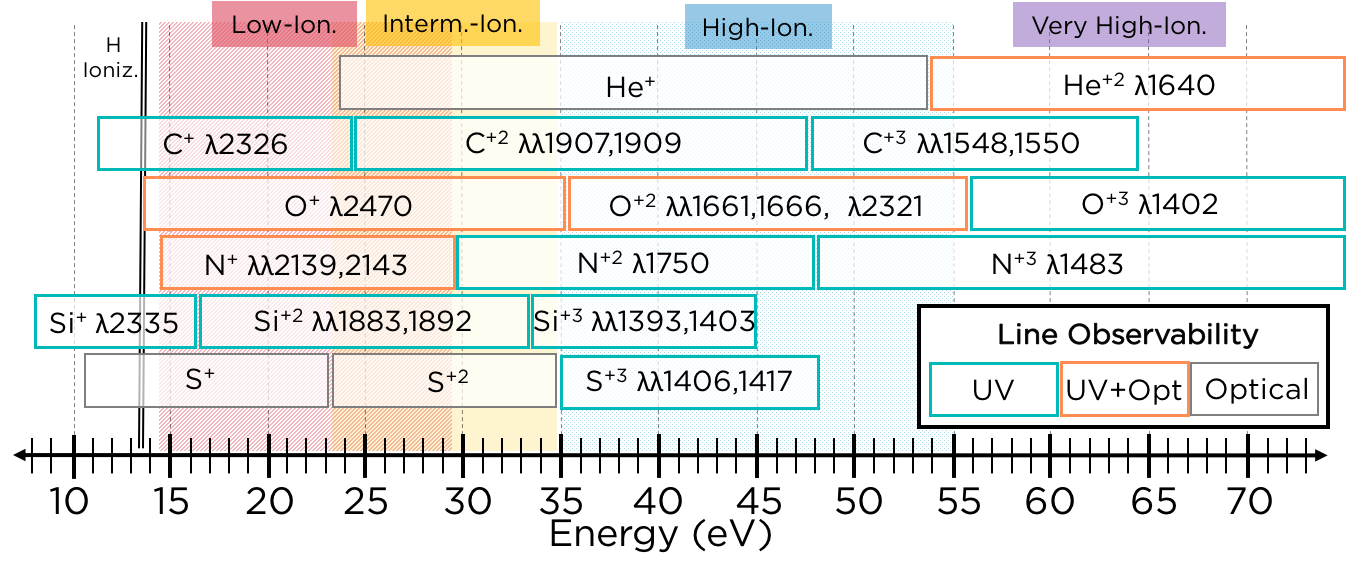} 
  \vspace{-2ex}
    \caption*{\scriptsize{Figure 2: Prominent emission-lines observed in nebulae. 
    Ions available in both the rest-frame UV and optical are outlined in orange, whereas the 
    additional ions afforded by the UV alone are outlined in green with their wavelengths listed. 
    Not only does the UV provide ions in common with the optical that will allow calibration of the UV
    diagnostics to the optical, but also greatly expands our ability to map complex ionization structures
    from low- to very high-ionization.}}
\vspace{-2ex}
\end{SCfigure}

\item There are many ongoing efforts to use nebular UV emission lines to constrain the radiation field within SFGs (e.g., Feltre+16; Jaskot+16; Byler+18; Nakajima+18). In particular, combinations of UV emission lines, such as C~{\sc iv}/He~\ii\ vs O~\iii]/He~\ii, may provide useful UV versions of the canonical ``BPT" diagnostic diagram. However, the UV and optical offer complimentary abilities to map the ionization structure of nebulae. 
The most prominent optical and UV nebular emission-line features are shown in Figure~2, where the UV
($1200-2500$ \AA) offers an expanded ionization energy window and ions not observable in the optical.
Spatially mapping the complex optical+UV ionization structure would provide great utility in constraining radiation hardness and the conditions that promote Ly$\alpha$ and LyC escape.
Overall, the rest-UV continuum offers a wealth of additional constraints over the optical regime. 
\end{itemize}

Multiple UV line detections across multiple sightlines currently exist in only a handful of high-$z$ gravitationally lensed objects (e.g., MEGaSaURA, Rigby+18).
The utility of all UV emission-line diagnostics is, therefore, severely limited at present. 
First, owing to this paucity of sources with multiple UV line detections, UV diagnostics have not been confirmed empirically, and completely lack the necessary spatial resolution testing.
Further, many of the strongest lines (C~\iv, He~\ii, Si~\iii], C~\iii]) are complicated by unknown contributions from nebular emission, stellar winds and photospheric absorption, ISM absorption, and dust (Leitherer+11).   
Opportunely, in the UV, the massive stars are responsible for producing both the underlying stellar continuum and the ionizing radiation that drives the nebular emission, providing further constraints on age, metallicity, and star-formation history and contributions to emission features.
Therefore, high spatial and spectral resolution UV emission-line mapping is needed to robustly model and disentangle the components of the high-ionization emission line diagnostics that are critical to the interpretation of distant sources.

\begin{figure}
  \centering
  \includegraphics[scale=0.525,trim=0mm 0mm 0mm 0mm,clip]{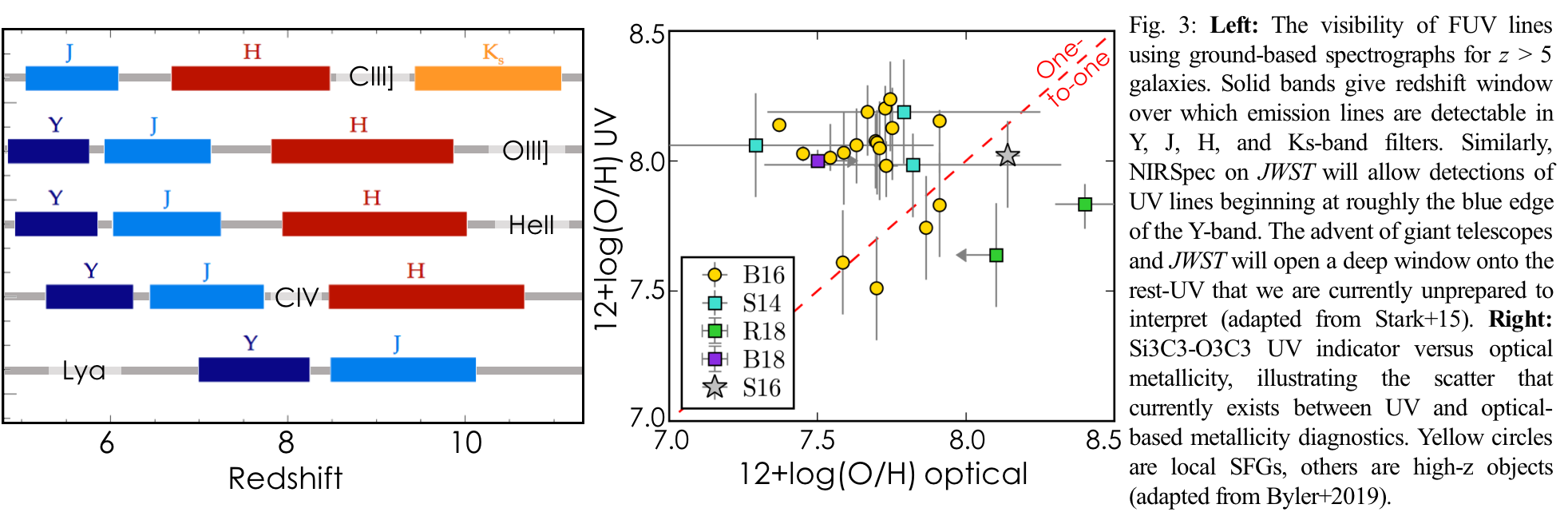} 
\vspace{-2ex}
\end{figure}

\section{Planning for the High-$z$ Era: The Need for a Space-Based rest-UV IFU}
\textit{One major issue currently remains in trying to constrain the UV diagnostics: we are not comparing UV and optical emission from the same H~\ii\ region.} We know from spatially resolved optical studies that integrated light masks a multitude of different chemical and physical conditions (e.g., Figure~1) and line ratios from this mixed light can lead to incorrect results. Many of the current attempts to connect the UV and optical emission involve integrated spectra across the source. \textit{In order to truly understand rest-frame UV diagnostics we need to spatially resolve the UV emission lines across nearby SFGs.} The calibration of such diagnostics can then be achieved via complementary spatially resolved observations with ground-based optical IFUs (e.g., MUSE or KCWI).
Our current access to the UV, however, is limited to local point-source or integrated-light studies with the {\it HST}, where mapping several positions is extremely costly and inefficient (i.e., with {\it HST}/COS, Fig.~1). Spatially resolved rest-frame UV studies of high-$z$ galaxies with rest-frame optical instruments can be achieved (e.g. Fig.~1) but such targets are either extremely faint or are gravitationally lensed. While the latter offers much-needed magnification, large uncertainties in 2D flux distributions often exist due to complicated lens modeling. Moreover, such objects simply cannot provide the high (H~\ii\ region-scale) spatial resolution required. \looseness=-2

The science goals outlined here require a spectroscopic instrument with a wavelength range of $\sim1200-2500$~\AA, covering a wealth of UV nebular emission, ISM absorption, and stellar lines.
$R > 5000$ is needed to match that of our best stellar continuum models ($\Delta \lambda = 0.4$~\AA\ for Starburst99) and disentangle features with multiple confounding components. 
Moreover, efficient mapping of rest-frame UV lines across nearby SFGs requires the combined power of a 10\,m class space-based telescope (e.g., LUVOIR) with a high throughput IFU or mutli-object spectrometer (MOS).
IFUs allow for the continuous spatial mapping at higher resolution, thus removing any complications due to integrated light and/or slit losses when determining line ratios. Spatial resolution on the order of 0.5\arcsec\ would be optimal to match the optical performance of MUSE with adaptive optics. \looseness=-2

The instrument would observe the plethora of nearby SFGs that are deemed reliable analogues to those seen at high-redshift: low metallicity (12+log(O/H) $<8.5$), high star-formation rate ($\gtrsim1 \ M_{\odot}\ {\rm yr}^{-1}$), and low mass ($M \lesssim10^{10}\ M_{\odot}$). Their nearby nature ($D<20$~Mpc) allows for high spatial resolution such that H~\ii\ regions ($\sim100$~pc in size) could be explored fully, and shock fronts (typically $\sim$10~pc in size) would be resolved in the nearest targets. Most crucially is that a wealth of ancillary rest-optical data that already exists for these objects (unlike their high-$z$ counterparts): 
spatially-resolved optical spectra from IFUs such as VLT/MUSE and Keck/KCWI, 
as well as high-resolution imaging of their stellar populations from HST/WFC3 and ACS.
The scientific application for an observational set-up of this kind is far-reaching. 
Further science cases would include the mapping of outflows in nearby objects, kinematical mapping of Ly$\alpha$ emission, spatial mapping of LyC escape (at $z>0.3$), and even the mapping of CGM in the halos of galaxies.
\textit{\textbf{All that prevents us from fully harnessing the power of UV diagnostics and exploiting the ground-breaking observations of the high-$z$ galaxy era, is the complementary spatially resolved UV information.}}

\clearpage

\bibliographystyle{plain}
\bibliography{references}
Baldwin, J.~A., et al. 1981, PASP, 93, 5 \\%
Barrera-Ballesteros, et al. 2018, ApJ, 852, 74\\%
Bayliss, M., et al. 2014, 790, 144 \\%
Belfiore, F., et al. 2016, MNRAS, 461, 3111 \\%
Berg, D., et al. 2015, ApJ, 806 \\%
Berg, D., et al. 2018, ApJ, 859, 2 \\%
Berg, D., et al. 2019, accepted to ApJ \\ %
Bik, A., et al. 2018, A\&A, 619, 131\\ %
Byler, N., et al. 2018, ApJ, 863, 14 \\%
Byler, N., et al. 2019, \textit{in-prep} \\ %
Christensen, L., et al. 2012, MNRAS, 427, 1973 \\ %
Curti, M., et al. 2017, MNRAS, 465, 1384 \\ %
D'Agostino, J. J., et al. 2019, arXiv:1902.10295\\ %
Davies, R. L., et al. 2017, MNRAS, 470, 4974 \\ %
D\'{i}az, et al. 2000, MNRAS, 318, 462 \\ %
Du, X., et al. 2017, ApJ, 838, 63 \\%
Erb, D. K., et al. 2010, ApJ, 719, 1168 \\
Feltre, A., et al. 2016, MNRAS, 456, 3354 \\%
Fensch, J., et al. 2016, 585, 17\\ %
Groves, B., et al. 2008, ApJS, 176, 438 \\ %
Ho, I.-T., et al. 2015, MNRAS, 448, 2030 \\ %
James, B. L., et al. 2014, MNRAS, 440, 1794 \\%
James, B. L., et al. 2016, ApJ, 816, 40 \\%
James, B. L., et al. 2018, MNRAS, 476, 1726 \\ %
James, B. L. et al. 2009, MNRAS, 398, 2 \\%
James, B. L., et al. 2013a, MNRAS, 430, 2097 \\%
James, B. L., et al. 2013b, MNRAS, 432, 2731 \\%
Jaskot, A. \& Ravindranath, S., 2016, ApJ, 833, 136 \\%
Jaskot, A., et al. 2017, ApJL, 851, L9 \\%
Jaskot, A., et al. 2019, in-prep \\  %
Kauffmann, G., et al. 2003, MNRAS, 346, 1055 \\%
Keenan, R. P., et al. 2017, MNRAS, 848, 12 \\ %
Kewley, L. J., et al. 2001, ApJ, 556, 121 \\%
Kewley, L. J., \& Dopita, M. A., 2002, ApJSupp., 142, 35\\ %
Kewley, L. J., et al. 2013, ApJ, 774, 100 \\%
Kobulnicky, H. A. \& Kewley, L. J., 2004, ApJ, 617, 240 \\ %
Kumari, N., et al. 2017, MNRAS, 470, 4618 \\%
Kumari, N., et al. 2018, MNRAS, 476, 3793 \\%
Lamareille, F., 2010, A\&A, 509, A53 \\%
Leitherer, C., et al. 2011, AJ, 141, 37 \\%
Maiolino, R. \& Mannucci, F., 2019, A\&ARv, 27, 3 \\ %
Micheva, G., et al. 2018, ApJ, 867, 2M \\ %
Micheva, G., et al. 2019, preprint, (arXiv:1902.03952) \\ %
Nagao, T., et al. 2006, A\&A, 459, 85 \\%
Nakajima, et al. 2018, A\&A, 612, 94 \\%
Newman, S. F., et al. 2014, ApJ, 781, 21 \\ %
Patricio, V., et al. 2016, MNRAS, 456, 4191 \\ %
Peimbert, M., 1967, ApJ, 150, 825 \\ %
Pilyugin, L. S. \& Thuan, T. X., 2005, ApJ, 631, 231 \\ %
Poetrodjojo, H., et al. 2018, MNRAS, 479, 5235 \\ %
Rigby, J., et al. 2015, ApJL, 814, L6 \\%
Rigby, J., et al. 2018, AJ, 155, 10 \\ %
Rich, J. A., et al. 2011, ApJ, 734, 87 \\ %
Sanders, R., et al. 2016, ApJ, 816, 23 \\
Sanders, R., et al. 2017, ApJ, 850, 136\\%
S\'{a}nchez, et al. 2014, A\&A, 563, A49 \\ %
Searle, L., 1971, ApJ, 168, 327 \\%
Senchyna, P., et al. 2017, MNRAS, 472, 2608 \\%
Shapley, A., et al. 2003, ApJ, 588, 65 \\ %
Stark, D., et al. 2015, MNRAS, 450, 1846 \\
Stark, D., et al. 2017, MNRAS, 464, 469\\%
Westmoquette, M. S., et al. 2013, A\&A, 550, A88 \\%
Wielbacher, P., et al. 2018, A\&A, 611, 95 \\ %
Zastrow, J., et al. 2011, ApJ, 741, 17\\ %

\end{document}